# Spin control with triplet and doublet excitons in organic semiconductors

Qinying Gu,[a] ‡ Sebastian Gorgon,[a] ‡ Alexander S. Romanov,[b] Feng Li,[c] Richard H. Friend[a],* and Emrys Evans[d],*

[a] *Department of Physics, University of Cambridge, Cambridge, CB2 1EW, U.K.*

[b] *School of Chemistry, University of Manchester, Manchester, M13 9PL, UK.*

[c] *State Key Laboratory of Supramolecular Structure and Materials, College of Chemistry, Jilin University, Qianjin Avenue 2699, Changchun, 130012, P. R. China*

[d] *Department of Chemistry, Swansea University, Singleton Park, Swansea SA2 8PP, UK*

‡ These authors contributed equally

*Correspondence should be addressed to: rhf10@cam.ac.uk (RHF), emrys.evans@swansea.ac.uk (EWE).

**Abstract**

Spin triplet exciton formation sets limits on technologies using organic semiconductors that are confined to singlet-triplet photophysics. In contrast, excitations in the spin doublet manifold in organic radical semiconductors can show efficient luminescence. Here we explore the dynamics of the spin allowed process of intermolecular energy transfer from triplet to doublet excitons. We employ a carbene-metal-amide (CMA-$CF_3$) as a model triplet donor host, since following photoexcitation it undergoes extremely fast intersystem crossing to set up a population of triplet excitons within 4 ps. This enables a foundational study for tracking energy transfer from triplets to a model radical semiconductor, TTM-3PCz. Over 90% of all radical luminescence originates from the triplet channel in this system under photoexcitation. We find that intermolecular triplet-to-doublet energy transfer can occur directly and rapidly, with 12% of triplet excitons transferring already on sub-ns timescales. This enhanced triplet harvesting mechanism is utilised in efficient near-infrared organic light-emitting diodes, which can be extended to other opto-electronic and -spintronic technologies by radical-based spin control in molecular semiconductors.

**Introduction**

Triplet excitons play a critical role in all technologies based on organic semiconductors, and effective spin management is required to achieve the desired optoelectronic properties. For closed-shell materials, organic light-emitting diodes (OLEDs) have well-known efficiency limits from the formation of 25% singlet and 75% triplet excitons with spin-allowed and -forbidden emission. The generation of dark triplet excitons that competes with charge extraction also hinders the performance in organic photovoltaics (OPV). The control of singlet and triplet states is critical in up- and down-conversion for photon management in imaging and increasing the spectral working range of devices such as photodetectors. In these technology areas it is necessary to obtain light emission from triplet excitons, which can be achieved by a spin-flip (e.g. phosphorescence, thermally activated delayed fluorescence (TADF), triplet-triplet annihilation[1,2]) with performance limits set by singlet-triplet photophysics. Here we show the advantage of going beyond the singlet-triplet manifold in the presence of an open-shell organic radical with an all-doublet spin energy landscape. We present a foundational study of the triplet-doublet control mechanisms for harvesting energy from dark triplets with effective use as doublet excitons on organic radicals for luminescence.

To demonstrate the potential of triplet-doublet photophysics we designed a model system for studying direct spin-allowed energy transfer from triplet excitons generated within a closed-shell organic host to a doublet radical chromophore. We use a carbene-metal-amide (CMA-CF$_3$, Figure 1a) as a model host since following photoexcitation it undergoes extremely fast picosecond intersystem crossing (ISC) to generate a population of triplet excitons. This enables a direct measurement by optical spectroscopy, where the clock starts from picoseconds for tracking the subsequent energy transfer from triplets to TTM-3PCz (Figure 1a) radical doublet states for light emission. We show one application of this where our photophysical studies reflect the working mechanism in an efficient near-infrared OLED (705 nm, > 16% external quantum efficiency) for which the core process that needs to happen fast and efficiently is transfer from triplet to doublets for electroluminescence. A rapid spin-allowed triplet-doublet transfer process overcomes the rate and performance limiting spin-flip step for light emission from triplet excitons in organic semiconductors.

Related singlet exciton energy transfer from a host TADF OLED system to a narrow band guest emitter (so-called hyperfluorescence) has been engineered to provide promising blue OLEDs[3,4]. The spin doublet emitter allows scope for direct energy transfer from the triplet excitons present in a host since this energy transfer process can be spin allowed. This was shown for a 4CzIPN (1,2,3,5-tetrakis(carbazol-9-yl)-4,6-dicyanobenzene) triplet harvesting host[5] and in the reverse process of doublet to triplet transfer[6] and in solution[7], although these demonstrations were limited to microsecond timescales. Direct triplet energy transfer is attractive since it can speed up the emission rate by bypassing the slow reverse ISC step involving a spin-flip in TADF.

The host CMA-CF$_3$ is chosen from a family of compounds with a molecular design of linear, two-coordinated coinage metal complexes where a metal atom bridges a cyclic (alkyl)(amino)carbene and carbazole as tunable electron π-acceptor and -donor components[8-24]. The generally observed picosecond ISC in carbene-metal-amides is driven by strong spin-orbit coupling derived from the coinage metal. Triplet states in carbene-metal-amides are long-lived up to microsecond timescales. By having tunable electronic and photophysical properties – whilst maintaining rapid ISC – as well as high thermal stability and insensitivity to aggregation quenching, this makes carbene-metal-amides the ideal host in model host-guest systems with organic radicals to study the mechanisms and fastest kinetics for triplet-doublet photophysics[25].

The guest TTM-3PCz in our model system is a luminescent π-radical emitter that operates within a doublet spin energy manifold. The optical and doublet-spin properties of radicals enable efficient OLED demonstrations using a mechanism that avoids detrimental triplet exciton formation in device action. Despite highly efficient OLED performance being reported for TTM-3PCz (Figure 1a) with 703 nm emission[26], charge imbalance and accumulation at interface layers in this design is found to limit the device efficiency, stability and cause substantial reduction in performance at high current density ('roll-off').[27] Here the triplet-doublet photophysical mechanisms revealed from our studies of CMA-CF$_3$:TTM-3PCz enables enhanced performance for radical OLEDs. Triplets formed by direct electrical excitation of host sites, away from the radical, are harvested via energy transfer to the doublet emitter in an application where spin-flip limits can be removed in emission pathways starting from triplet excitons. We show that spin-allowed triplet-doublet energy transfer is efficient already on sub-nanosecond timescales, which is orders of magnitude faster than in triplet management technologies using singlet-triplet-limited photophysics[1,2].

**Results and Discussion**

For energy transfer we require that the HOMO and SOMO levels of the TTM-3PCz energy acceptor lie within the HOMO and LUMO levels of the carbene-metal-amide host. Energy level values for TTM-3PCz and CMA-CF3 were previously obtained from cyclic voltammetry[10,26]: HOMO$_{TTM-3PCz}$ = −6.0 eV; SOMO$_{TTM-3PCz(reduction)}$ = −3.7 eV; HOMO$_{CMA-CF3}$ = −6.12 eV; LUMO$_{CMA-CF3}$ = −2.92 eV. We have tried several host candidates and found it was necessary to use two trifluoromethyl groups on the carbazole moiety in order to lower the HOMO energy of CMA-CF$_3$ below the TTM-3PCz HOMO to set up conditions for excited-state energy transfer to the radical component (Figure 1a). Other CMA candidates with cyano substitution and single trifluoromethyl have HOMO energies of −5.80 eV and higher, and led to substantial host CMA emission in photoluminescence and electroluminescence from CMA:TTM-3PCz combinations as presented in Fig. S1. Favourable pairing of CMA-CF$_3$ and TTM-3PCz for energy transfer is demonstrated in Figure 1b where optical properties in tetrahydrofuran (THF) solvent show a significant spectral overlap of CMA-CF$_3$ delayed fluorescence (dark blue, measured at room temperature) and phosphorescence (light blue, measured at 77K) with the TTM-3PCz absorption profile (red). This supports energy conditions for transfer of excitations from singlet and triplet exciton states of CMA-CF$_3$ to TTM-3PCz radical. CMA-CF$_3$ shows structured deep blue emission ($\lambda_{max}$= 425 nm) in frozen 2-MeTHF at 77 K which is ascribed to local (i.e. ligand-centred) excited luminescence The emission red-shifts in liquid toluene solution at room temperature and becomes broad and unstructured in this less-constrained environment, and is assign to CT excited states[10] (Figure S2 for transient PL spectrum of solid-state CMA-CF$_3$). A broad and structured photoluminescence (PL) profile (Figure 1c) with major peaks at 425 nm and 470 nm is measured for the pristine CMA-CF$_3$ film, indicating that the emission in the solid state is mainly from the local-excited triplet state from the carbazole ligand, $^3$LE(Cz), and is mixed with delayed luminescence of charge transfer CT character. Therefore, unlike the standard CMA1[28], the emission of CMA-CF$_3$ films shows substantial phosphorescence contribution[10]. Thin-film PL of TTM-3PCz shows a strong CT character ascribed to electronic transitions from TTM-centred SOMO to 3PCz-HOMO in doublet fluorescence from D$_1$ (707 nm emission peak)[26].

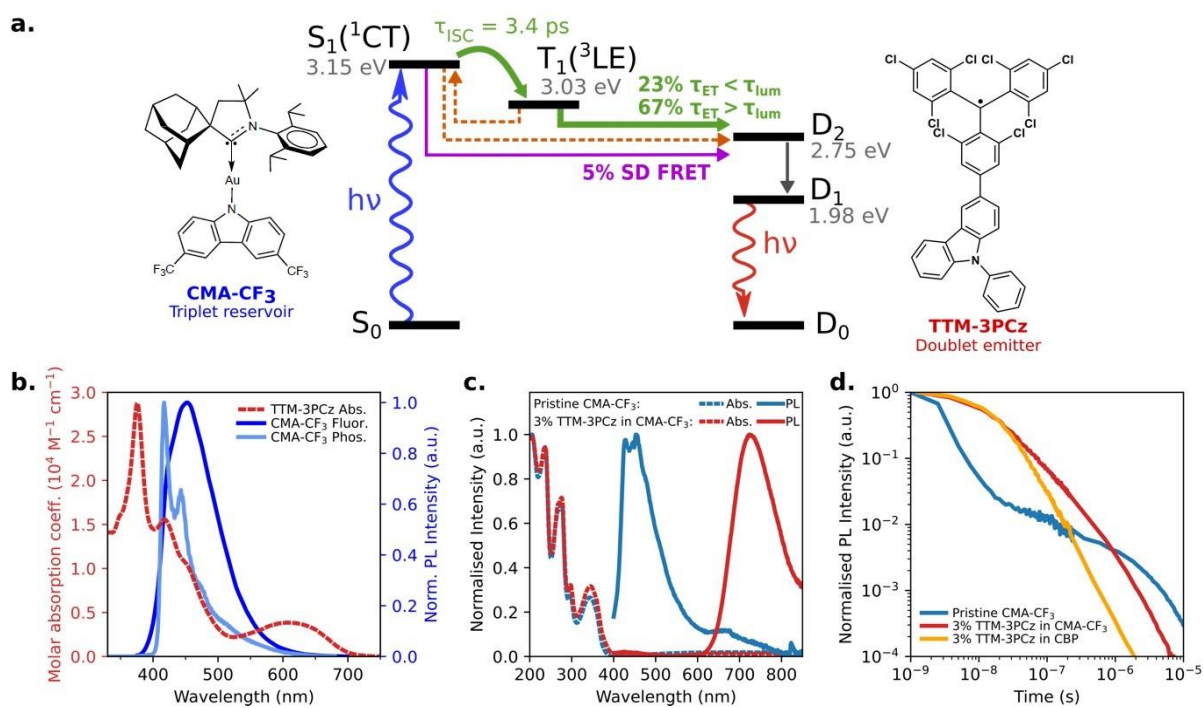

Figure 1. **Model triplet-doublet energy transfer system with CMA-CF$_3$ and TTM-3PCz**. (a) Chemical structures of CMA-CF$_3$ and TTM-3PCz, and energy transfer mechanism in 3% TTM-3PCz in CMA-CF$_3$ blend with contributions to TTM-3PCz emission indicated as percentages. (b) Normalised molar absorption coefficient (red dashed line) of TTM-3PCz in THF, fluorescence (dark blue line) and phosphorescence (light blue line) spectra of CMA-CF$_3$ liquid toluene solution and frozen MeTHF respectively (excitation at 365 nm). Significant spectral overlap of the absorption and emission profiles enables energy transfer from $S_1$ and $T_1$ of CMA-CF$_3$ to TTM-3PCz. (c) Absorption and steady-state PL spectra of 3% TTM-3PCz in CMA-CF$_3$ and pristine CMA-CF$_3$ films. CMA-CF$_3$ emission is completely quenched in the blend with only radical emission observed. (d) Time-resolved PL of 3% TTM-3PCz in CMA-CF$_3$, 3% TTM-3PCz in 4,4'-bis (N-carbazolyl)-1,1'-biphenyl (CBP) and pristine CMA-CF$_3$ films. Delayed emission in 3% TTM-3PCz in CMA-CF$_3$ shows an activated energy transfer channel for radical emission that is faster than pristine CMA-CF$_3$, but slower than 3% TTM-3PCz in CBP.

To examine the effectiveness of CMA-CF$_3$ as a host for energy transfer, we measured the absorbance and steady-state PL of the pristine CMA-CF$_3$ film and 3% TTM-3PCz in CMA-CF$_3$ blend (Figure 1c) at film thickness of 80 nm. We excited the films at 355 nm, where CMA-CF$_3$ is strongly absorbing. The blend exhibits PL spectra near 715 nm due entirely to TTM-3PCz, and the CMA-CF$_3$ emission is quenched below the noise floor. Here TTM-3PCz emission is red-shifted by ca. 10 nm compared to films in CBP, which is attributed to host polarity effects (Figure S3). As direct TTM-3PCz photoexcitation is negligible, this shows that the radicals efficiently harvest CMA-CF$_3$ excitons by energy transfer for doublet emission.

**Time-resolved spectroscopy**

For tracking the dynamics of energy transfer, we performed time-resolved PL on ns-µs timescales. A 3% TTM-3PCz in CBP film was used as a reference, since CBP has been shown to be a successful host in achieving efficient radical emission[26]. All samples were excited at 355 nm with fluences near 5 µJ/cm$^2$. As shown in Figure 1d, pristine CMA-CF$_3$ shows a multiexponential decay with prompt fluorescence from $^1$CT on timescales shorter than the system time resolution (2 ns), followed by mixture of delayed fluorescence and phosphorescence from $^3$CT and $^3$LE as previously reported[10]. However, on pairing CMA-CF$_3$ with TTM-3PCz, the 3% TTM:3PCz in CMA-CF$_3$ PL kinetics approaches that of the 3% TTM:3PCz in CBP blend. PL time slices exhibit only deep-red emission from the TTM-3PCz radical for the duration of the emission lifetime (Figure S4a). The delayed component of the radical emission in the model blend is faster than delayed host emission in pristine CMA-CF$_3$, but slower than radical emission in 3% TTM-3PCz in CBP, suggesting that triplets are involved in energy transfer for TTM-3PCz in CMA-CF$_3$ blends.

To establish the earlier time kinetics, we performed transient absorption (TA) spectroscopy on ps-µs timescales. Following 355 nm excitation to achieve majority CMA-CF$_3$ host excitation, several overlapping photoinduced absorption (PIA) bands are present in the 3% TTM-3PCz in CMA-CF$_3$ blend, as shown in Figure 2a. To assign these excited-state features, we performed TA experiments on reference samples (Figure S5). Photoexcitation of pristine CMA-CF$_3$ yields a broad PIA band centred at 660 nm that decays on ps timescales with formation of a narrower band at 590 nm, which we assign to the CMA-CF$_3$ singlet (S$_1$) and triplet (T$_1$) excitons respectively. To establish the D$_1$ exciton profile of TTM-3PCz, we used samples of 3% TTM-3PCz in CBP and observed two PIAs centred at 620 nm and 1650 nm, matching our previous solution studies[26,29].

As shown in Figure 2b, the bands identified in the control samples appear without any spectral shifts in the model 3% TTM-3PCz in CMA-CF$_3$ blend. Following photoexcitation into the CMA-CF$_3$ S$_1$ level, intersystem crossing (ISC) occurs within the first few picoseconds to yield T$_1$ excitons. The TTM-3PCz D$_1$ level population grows continually until nanosecond timescales. We used a genetic algorithm to extract population-specific kinetic traces for the visible probe region (SI: Extracting population specific dynamics from TA, p.7). The optimised species-associated spectra matched the bands seen in control samples (Figure S5a) and allowed S$_1$ and T$_1$ kinetic traces to be extracted from the visible probe region data. We use the kinetics from the 1650 nm IR PIA to extract D$_1$ kinetic traces. These are shown in Figure 2c for the 3% TTM-3PCz in CMA-CF$_3$ blend of interest, and two control samples of pristine CMA-CF$_3$ and 3% TTM-3PCz in CBP blend.

The S$_1$ dynamics for CMA-CF$_3$ are unchanged in the presence of TTM-3PCz, as both pristine host and 3% blend show rapid ISC with a lifetime of 3.4 ps. This indicates that the introduction of radical dopants does not affect ISC dynamics within CMA-CF$_3$ excited states and confirms that triplet excitons are in excess of singlet excitons in the model blend already from 2.4 ± 0.3 ps.

The D$_1$ signal in 3% TTM-3PCz in CMA-CF$_3$ blend peaked at 1.1 ns, two orders of magnitude later than in the 3% TTM-3PCz in CBP control sample. Both samples show comparable direct radical absorption at the excitation wavelength and similar early time dynamics. Rapid energy transfer from CBP occurs exclusively via Förster resonance energy transfer (FRET) from CBP singlets with a lifetime of 11.6 ps (Figure S5b), which is similar to the S$_1$-D$_1$ FRET lifetime for the 3% TTM-3PCz in CMA-CF$_3$ blend calculated as 12 ps (SI p.6). Therefore, in the TTM-3PCz in CMA-CF$_3$ blend, singlet to doublet FRET is largely outcompeted by rapid ISC, and occurs with a quantum yield of approximately 5%. Since singlet-doublet FRET is not the dominant energy transfer pathway, we can focus on the triplet-doublet energy transfer (TDET) channel contributions to the blend photophysics.

The T$_1$ population peaked at 38 ps in pristine CMA-CF$_3$, but at 24 ps in the 3% TTM-3PCz in CMA-CF$_3$ blend. It decayed faster with increasing TTM-3PCz loading, with its fastest decay lifetime decreasing from 200 ns in pristine CMA-CF$_3$, to 70 ns in 3%, and further to 30 ns in 5% TTM-3PCz in CMA-CF$_3$ blends (Figure S6, S7). This suggests that CMA-CF$_3$ triplet excitons undergo energy transfer to TTM-3PCz sites on picosecond to nanosecond timescales. We consider that the slower transfer rates are limited by the speed of triplet exciton hopping between CMA-CF$_3$ sites until they are in spatial range of a TTM-3PCz molecule, and forming an encounter pair in a spin configuration that allows energy transfer[30]. To estimate the fraction of TDET by a given time we compare the loss of $T_1$ population seen in TA of the blend versus pristine CMA-CF$_3$. We observe that 23% of triplet excitons transfer within one radical radiative lifetime and 12% of triplet excitons transfer within 1 nanosecond.

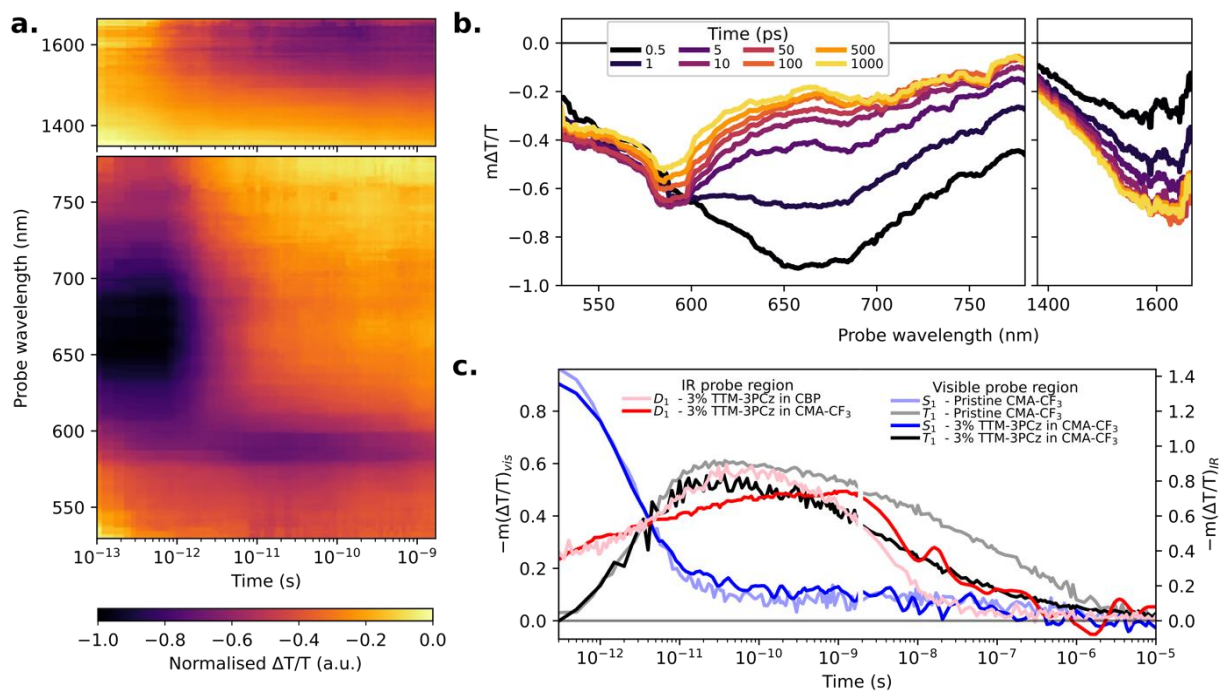

Figure 2. **Excited state dynamics of TTM-3PCz and CMA-CF$_3$.** (a) Normalised transient absorption spectroscopy map of 3% TTM-3PCz in CMA-CF$_3$ following 355 nm excitation at fluence of 5.1 μJ/cm$^2$, with (b) time slices of unnormalized spectra. PIA bands corresponding to CMA-CF$_3$ singlet $S_1$ and triplet $T_1$ excitons are centred at 660 nm and 590 nm respectively. TTM-3PCz doublet $D_1$ exciton PIAs are present at 1600 nm and 620 nm. (c) Kinetics extracted from TA in 3% TTM-3PCz in CMA-CF$_3$ blend the reference 3% TTM-3PCz in CBP blend and pristine CMA-CF$_3$. Following fast ISC, the CMA-CF$_3$ $T_1$ decays faster in presence of TTM-3PCz. The $D_1$ population peaks around 1 ns and tracks the decay of the $T_1$ signal for the 3% TTM-3PCz in CMA-CF$_3$ blend, compared to peaking at 60 ps and displaying a faster decay for the 3% TTM-3PCz in CBP blend.

The TA results show that triplet excitons play the dominant role in energy transfer. To gain insight into the mechanism of TDET we performed temperature-dependent time-resolved PL spectroscopy. At room temperature, analysis of the integrated PL (Figure 3a, inset) reveals that 24% of all emission in the blend occurs within one radical radiative lifetime. This matches the TA results and suggests the majority of triplet excitons transfer at later times. From the fraction of delayed radical emission, we estimate this fraction at 67%. Upon cooling, the delayed emission is significantly slower in both the 3% TTM-3PCz in CMA-CF$_3$ blend (Figure 3a) and pristine CMA-CF$_3$ (Figure 3b). TTM-3PCz in CMA-CF$_3$ showed no changes to the emission spectra upon cooling. We used Arrhenius analysis of $k_{RISC}$ (rate of reverse intersystem crossing, RISC) against the inverse of temperature to estimate the activation energies ($E_a$), following methodology for delayed fluorescence systems (SI p.8, Fig. S4). For pristine CMA-CF$_3$, the activation energy was found to be 70 ± 12 meV. The key temperature-activated process in pristine CMA-CF$_3$ is the RISC from triplet to singlet excitons, and this value shows good agreement with the calculated $\Delta E_{ST}$ for CMA-CF$_3$ (120 meV) from fluorescence and phosphorescence spectra. This large activation energy makes RISC relatively slow. However, for the 3% TTM-3PCz in CMA-CF$_3$ blend the activation energy calculated from the Arrhenius equation was found to be $E_a$ =24 ± 8 meV. This shows the delayed component of radical emission is not due to thermal excitation of triplet to singlet excitons followed by singlet-doublet energy transfer, hyperfluorescence, since this would require the same activation energy[25]. Therefore, as the activation energy is lower in blends compared to pristine CMA-CF$_3$, RISC to the singlet level is not mediating energy transfer from triplets to TTM-3PCz doublets. We assign this activation energy to triplet-diffusion limited reformation of triplet-doublet encounter pairs.

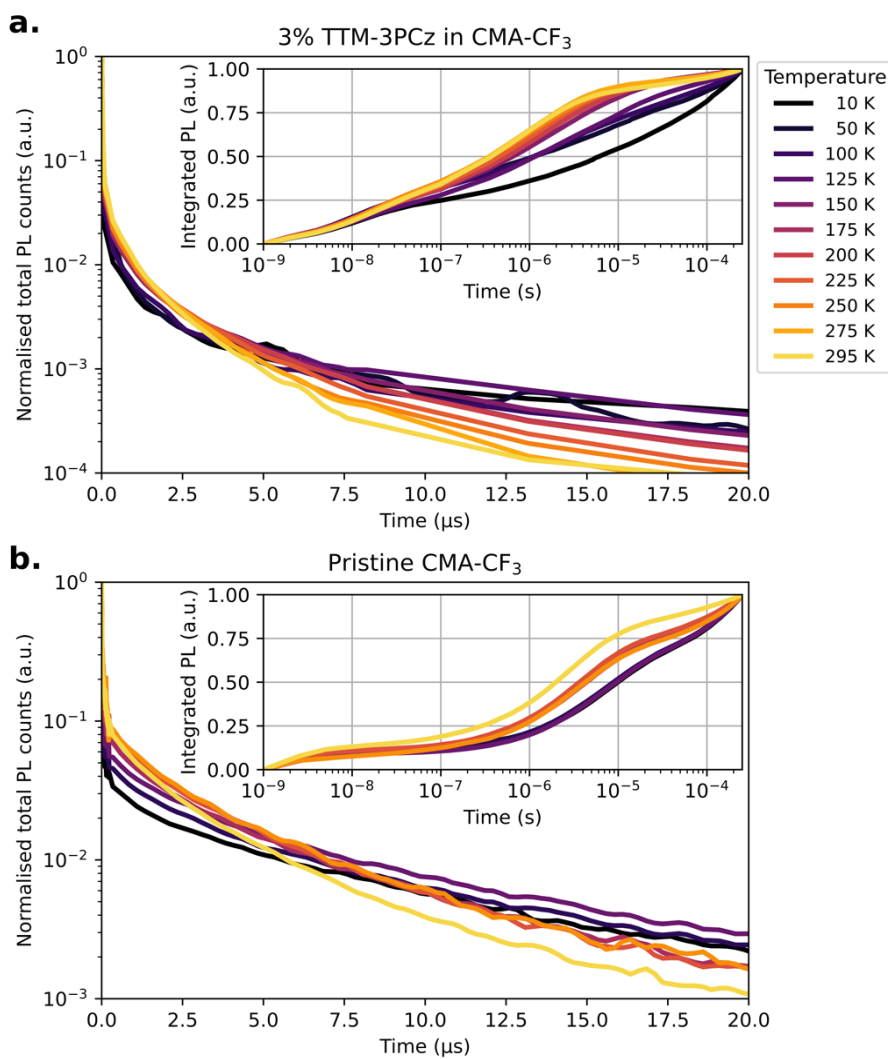

Figure 3. **Temperature-dependent emission dynamics of TTM-3PCz and CMA-CF$_3$.** Temperature-dependence on the time-resolved PL kinetics of (a) 3% TTM-3PCz in CMA-CF$_3$ and (b) pristine CMA-CF$_3$ films following 355 nm excitation. Insets show integrated PL. Lowering of activation energy compared to pristine host shows that hyperfluorescence is not responsible for delayed emission in 3% TTM-3PCz in CMA-CF$_3$ blend.

**Discussion of energy transfer mechanism**

Our time-resolved spectroscopy results demonstrate that T$_1$ triplet excitons accumulate rapidly following photoexcitation of CMA-CF$_3$ and this provides a model system to study the subsequent energy transfer from this triplet population. Due to the rapid ISC, SD FRET is not a significant channel in this blend. We observe energy transfer directly from triplets to form doublet excitons in TTM-3PCz, with some TDET occurring already on picosecond timescales. Two of the main energy transfer mechanisms in organic systems are: Förster resonance energy transfer (FRET) and Dexter energy transfer (DET)[30-40]. FRET occurs through a long-range dipole-dipole electromagnetic interaction and its efficiency is dependent on the spectral overlap integral between the emission profile of the host (CMA-CF$_3$) and the absorption of the dopant (TTM-3PCz)[37,41]. The triplets in CMA-CF$_3$ can also undergo FRET because CMA-CF$_3$ is phosphorescent. Although 'pure' triplets have no oscillator strength and are generally not active in FRET, enhanced spin-orbit coupling derived from the metal centre makes the CMA-CF$_3$ triplets emissive, and they can be radiatively dipole-coupled to radicals for energy transfer. FRET from triplet to singlet excitons has been reported in the literature for phosphorescent energy donors and fluorescent acceptors[42]. As shown in Figure 1b, the absorption of TTM-3PCz overlaps well with the phosphorescence spectrum of CMA-CF$_3$ indicating the triplet to doublet energy transfer via Förster type is possible. FRET can be therefore used to harvest energy from both singlet and triplet CMA-CF$_3$ excitons.

Energy transfer is also achievable via Dexter energy transfer (DET) in an electron-exchange mechanism, whose efficiency depends on the overlap of molecular orbitals on adjacent molecule. Unlike conventional fluorescent dopants, transfer from triplet to doublets can occur with spin conservation in 1/3 of the possible intermolecular encounter pairs[5].

Marcus theory has been applied in analysis of Dexter triplet-triplet energy transfer, where the reorganisation energy ($\lambda$) is found to slow isoenergetic energy transfer considerably[38]. We now apply it to the triplet-doublet case. For the system studied here, there is a large energy difference ($\Delta G^0$ =1.0 eV) between the T$_1$ and D$_1$ energy levels. We consider that this energy difference compensates the reorganisation energy required to attain a molecular geometry and environment for DET. This leads to a small Gibbs free energy of activation component (< 70 meV) for triplet-doublet transfer step to generate luminescent states, which is consistent with the PL temperature dependence of the system (see SI p.8-9, Fig. S8).

As the doublet population dynamics show a stretched rise and decay and DET requires close molecular proximity, we expect the picosecond-DET to occur from excited CMA-CF$_3$ sites directly adjacent to a TTM-3PCz, and the rate of nanosecond-DET to be limited by triplet diffusion (cf. Figure S7).

Additionally, part of the nanosecond energy transfer may originate in a FRET mechanism from distant T$_1$ excitons that have not encountered a TTM-3PCz site directly. As the phosphorescence of CMA-CF$_3$ and the absorption of TTM-3PCz overlap around 450 nm (Fig. 1b), which corresponds to the energy of the second doublet excited state (ca. 2.75 eV) of TTM-3PCz, direct FRET can take place from T$_1$ to D$_2$. D$_2$ excitons can undergo rapid internal conversion to D$_1$, and then fluorescence to the D$_0$ ground state.

In summary, we identify two direct triplet-doublet energy transfer routes: the dominant fast channel occurring on sub-20 ns times which accounts for 23% of emitted photons, made up of temperature independent Dexter transfer from T$_1$ to D$_1$ with a possible contribution of TD FRET for nanosecond timescales. This is followed by slower, temperature-activated triplet diffusion within the CMA-CF$_3$ host matrix where triplets can encounter a radical site in range and in an allowed spin configuration for TDET, leading to 67% of emitted photons. The smaller activation energy for delayed emission in the blend compared to the pristine host indicates that the excitation transfer process is direct and not mediated by 'hyperfluorescence'. The remaining emitted photons originate from excitons transferred via FRET from the singlet level at sub-10 ps times (5%) and weak direct TTM-3PCz absorption (5%). Our study pushes the boundary of emission times in organic semiconductors limited by delayed fluorescence or phosphorescence from triplet excitons[43]. We show triplet-doublet transfer can be an efficient and fast mechanism for harvesting emissive excitations in organic semiconductors using a luminescent radical energy sink where doublet excitons are degenerate or lower energy than triplet excitons in a system.

**Device performance and physics**

Our radical-based method for efficient light emission that originates from triplet excitons (Figure 1a) can advance performance and unlock possibilities for spin control in optoelectronics, optospintronics and photon management technologies from organic semiconductors. Here we target demonstrations in OLEDs to show one application that emerges from using triplet-doublet photophysics. Several device structures were prepared by vacuum thermal deposition and the following architecture was found to give the best performance: indium tin oxide (ITO) / molybdenum trioxide ($MoO_3$) (5nm) / 1,1-Bis[(di-4-tolylamino)phenyl]cyclohexane (TAPC) (40nm) / emissive layer (EML) (30 nm) / 4,6-Bis(3,5-di-3-pyridylphenyl)-2-methylpyrimidine (B3PYMPM) (60 nm) / lithium fluoride (LiF) (0.8 nm) / aluminium (Al) (100 nm) (Figure 4a). The optimised device shows near infrared (NIR) electroluminescence (EL, $\lambda_{max}$ = 705 nm) with a maximum external quantum efficiency (EQE) of 16.0 %, which ranks among the highest EQE values for NIR OLEDs employing purely organic emitters[44,45]. The low turn-on voltage (3.8 V) and high achievable radiance are also competitive compared to previously reported radical-based OLEDs[26,46]. The emission of the device is only from TTM-3PCz across the applied voltage from 5 V to 10 V (Figure 4d), with efficient quenching of the CMA-$CF_3$ EL. We optimised the OLED device stack in terms of the selection and thickness control of emissive layer (EML) and the electron transport layer (ETL) to achieve negligible current leakage and therefore good charge balance (see Figure S9-12) for current density–voltage and radiance–voltage characteristics of other device stacks). All devices show high EQEs at low current density together with good reproducibility.

The main exciton formation mechanism in 3% TTM-3PCz in CMA-$CF_3$ OLEDs is considered to be the energy transfer mechanism from excited CMA-$CF_3$ as studied in thin film photophysics. We also replaced the B3PYMPM for 2,2',2''-(1,3,5-Benzinetriyl)-tris(1-phenyl-1-H-benzimidazole) (TPBi): 1,4-Bis(triphenylsilyl)benzene (UGH2) (9:1 wt.%, 10 nm) and TPBi (60 nm) where TPBi and UGH2 are ETL and hole blocking layers, respectively. Here the LUMO energies are closer to that of CMA-CF$_3$ so that it is more preferentially excited over TTM-3PCz when using the TPBi:UGH2 transport layer compared to B3PYMPM[10,28]. This device shows 14.9 % maximum EQE and no residual emission from CMA-CF$_3$ can be seen (Figure S12), consistent with our energy transfer mechanism.

Additionally, magneto-electroluminescence (MEL) studies were conducted to obtain further mechanism insights from the carrier spin dynamics in working OLEDs[47-50]. In the reference 4,4′-Bis(N-carbazolyl)-1,1′-biphenyl (CBP)-based OLED (Figure S13), we found negligible MEL effects, which are consistent with previous observations in radical OLEDs with direct doublet electrical excitation[48]. A positive MEL was observed in the 3% TTM-3PCz in CMA-CF$_3$ device. The magnetosensitivity is attributed to Zeeman splitting[51,52] of triplet exciton sublevels, with positive MEL effects from increased triplet CMA-CF$_3$ exciton populations in the presence of magnetic field, and therefore an enhanced triplet-to-doublet energy transfer[53].

Device stability is a key property that must be realised for the practical use of radical OLEDs. Compared to 3% TTM-3PCz in CBP devices, where lifetimes are limited to minutes timescales, our device using the energy transfer mechanism has an order of magnitude of improvement in lifetime to around one hour at the same current density without device encapsulation (Figure S14 for the device stability comparison). We attribute the device improvement to the ability of CMA-CF$_3$ to promote electrical excitation away from radical sites in this design, with doublet excitons for infrared emission formed by efficient energy transfer from CMA-CF$_3$ to TTM-3PCz as demonstrated in our photophysics studies.

There is even larger improvement in device stability compared to the pristine CMA-CF$_3$. Degradation of carbene-metal-amides is considered to be due to the presence of excitons rather than charges[54]. This is particularly severe for the CMA-CF$_3$ material used here, for which device lifetimes are only a few seconds (Figure S14). However, in our device design, efficient energy transfer quickly depopulates the high-energy triplet excitons from CMA-CF$_3$, avoiding the degradation caused by high-energy triplets. Our successful demonstration of a radical energy transfer system in optoelectronics shows the potential of higher performance limits and novel technology platforms from singlet-triplet-doublet photophysics.

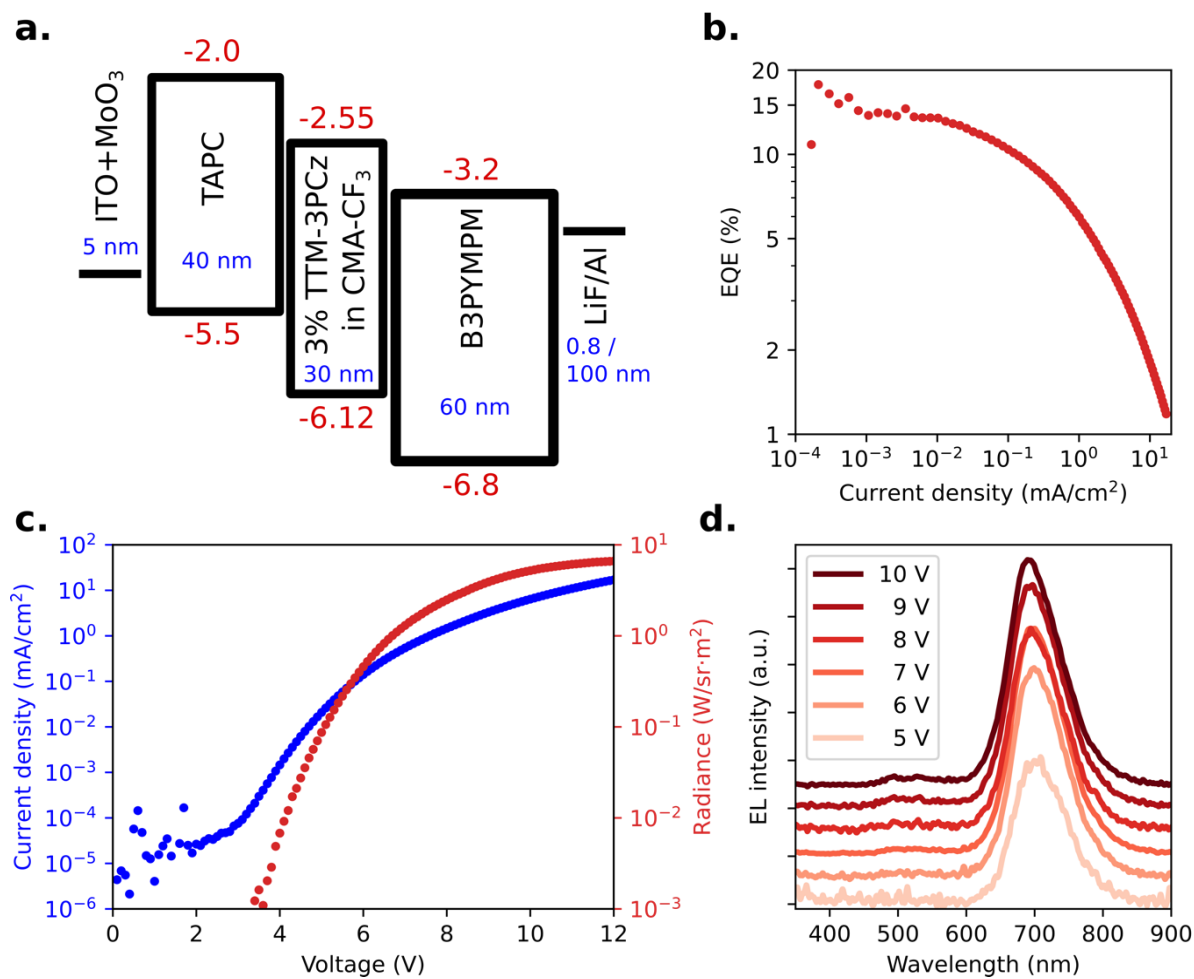

Figure 4. **OLED characteristics of TTM-3PCz and CMA-CF$_3$.** (a) Optimised device structure for 3% TTM-3PCz: CMA-CF$_3$ OLED. (b) EQE vs. current density with maximum EQE of 16.0 %. (c) Current density vs. voltage (blue) and radiance vs. voltage (red) of the 3% TTM-3PCz: CMA-CF$_3$ device. (d) EL spectra of the champion device from 5 V to 10 V indicating of 3% TTM-3PCz: CMA-CF$_3$ device is from TTM-3PCz.

**Conclusions**

In summary we demonstrated that the spin-allowed triplet to doublet energy transfer process can occur on sub-nanosecond timescales. These foundational studies of a triplet-doublet energy manifold were conducted in a model system using triplet excitons from CMA-CF$_3$ and doublet-spin radicals from TTM-3PCz. Due to the few-picosecond ISC in CMA-CF$_3$, which leads to rapid generation of triplet excitons in high yield by optical excitation, the triplet-doublet transfer mechanisms can be tracked in spectroscopic studies of the model system. We show that some excitons transfer from CMA-CF$_3$ T$_1$ to TTM-3PCz D$_1$ states within hundreds of picoseconds and attribute this to an efficient triplet-doublet process, showing that triplet management mechanisms for emission can be accelerated by orders of magnitude compared to spin-flip schemes in standard singlet-triplet photophysics. A demonstration of OLEDs using this system translated the triplet-doublet mechanism enabling emission from triplet excitons that otherwise limit the device performance. High EQEs of up to 16.0 % were achieved in TTM-3PCz:CMA-CF$_3$ devices with near infrared 705 nm electroluminescence derived from triplet-doublet energy transfer. The mechanism is applicable to all technology platforms using organic semiconductors in applications ranging from imaging to communications, to where radical spin control overcomes performance limits in opto-electronics and -spintronics, and the possibilities created from enhanced luminescence in molecular materials.

**Acknowledgements**

QG is grateful to the Cambridge Trust and China Scholarship Council (Grant No.201808060075) for the financial support. SG acknowledges funding from the EPSRC Centre for Doctoral Training in Integrated Functional Nano (Grant EP/S022953/1) and Christ's College, Cambridge. FL is grateful for financial support from the National Natural Science Foundation of China (grant no. 51925303) and the programme 'JLUSTIRT' (grant no. 2019TD-33). FL is an academic visitor at the Cavendish Laboratory, Cambridge, and is supported by the Talents Cultivation Programme (Jilin University, China). RHF acknowledges support from the Simons Foundation (grant no. 601946) and the EPSRC (EP/M005143/1). EWE is grateful to the Leverhulme Trust for an Early Career Fellowship; and Royal Society for a University Research Fellowship (grant no. URF\R1\201300). This project has received funding from the ERC under the European Union's Horizon 2020 research and innovation programme (grant agreement number 101020167).

**Author Contributions**

QG fabricated thin-films and optimised OLED devices, which were characterised by photoluminescence, device performance measurements and magnetic field studies. SG carried out the transient absorption measurements, data analysis and model calculation. QG and SG performed the temperature-dependent time-resolved PL. FL and AR synthesised and provided the materials. EWE and RHF conceived the project and supervised the work. All the authors contribute to results analysis and manuscript writing.

**Additional information**

Supplementary information accompanies this paper at: [accession code to be completed in proofs]

**Competing financial interests**

The authors declare no competing interests.

**Data availability**

The data underlying this article are available at: [accession code to be completed in proofs]

**Code availability**

Code used to analyse data in this manuscript are available from the corresponding authors upon reasonable request.

**Methods**

**Materials**
TTM-3PCz and CMA-CF$_3$ were synthesised as previously reported[10,26]. TAPC, B3PYMPM, CBP of sublimed grade and other OLED materials were obtained from Ossila and Lumtec.

**Photophysics**
Optical spectroscopy measurements (time-resolved PL and transient absorption, TA) were conducted on home-built setups powered by a Ti:Sapphire amplifier (Spectra Physics Solstice Ace, 100 fs pulses at 800 nm, 7 W at 1 kHz). Time-resolved PL studies were performed with an Andor spectrometer setup using a spectrograph and electrically gated intensified CCD camera (Andor SR303i; Andor iStar). Sample photoexcitation with 355 nm pump pulses were provided by frequency doubling of the output of a home-built narrowband visible Noncollinear Optical Parametric Amplifier (NOPA) pumped by 1 W of the Ti:Sapphire output for time-resolved PL experiments. 355 nm pump pulses for the short-time (<2 ns) TA studies were prepared in a similar way. The third harmonic of an electronically triggered Q-switched Nd:YVO4 (1 ns pump length, Advanced Optical Technologies Ltd AOT-YVO-25QSPX) laser was used as the source of 355 nm pump pulses for the long-time (> 1ns) TA studies. Broadband NOPAs were used to generate probe pulses for TA in the visible (510-780 nm) and infrared (1250-1650 nm) wavelength ranges. NOPA probe pulses were divided into two identical beams by a 50/50 beamsplitter, allowing the use of a reference beam for improved signal:noise. Custom-built Si (Hamamatsu S8381-1024Q) and InGaAs (Hamamatsu G11608-512DA) dual-line arrays from Stresing Entwicklungsbüro were used to detect probe and reference pulses.

**Device fabrication and characterisation**

An Angstrom Engineering EvoVac 700 system was used to fabricate organic semiconductor films and devices by vacuum deposition ($10^{-7}$ torr). Current density, voltage, electroluminescence characteristics and device lifetimes were obtained from a Keithley 2400 sourcemeter, Keithley 2000 multimeter and calibrated silicon photodiode in a home-built setup. Magneto-electroluminescent (EL) measurements were performed with Andor spectrometer (Shamrock 303i and iDus camera) to monitor modulation of EL in presence of magnetic field applied by GMW 3470 electromagnet.